\documentclass[english,reprint,aps,pra,showpacs,superscriptaddress,twocolumns]{revtex4-1}
\usepackage[T1]{fontenc}
\usepackage[latin9]{inputenc}
\usepackage{color}
\usepackage{babel}
\usepackage{textcomp}
\usepackage{mathrsfs}
\usepackage{amsmath}
\usepackage{amssymb}
\usepackage{graphicx}
\usepackage{esint}
\usepackage[unicode=true,pdfusetitle,
 bookmarks=true,bookmarksnumbered=false,bookmarksopen=false,
 breaklinks=true,pdfborder={0 0 0},backref=false,colorlinks=true]
 {hyperref}

\makeatletter
\usepackage{lipsum}

\makeatother

\begin{document}
\title{Dimerized Decomposition of Quantum Evolution on an Arbitrary Graph}

\author{He Feng}
	\affiliation{Shanghai Advanced Research Institute, Chinese Academy of Sciences, Shanghai 201210, China}
	\affiliation{University of Chinese Academy of Sciences, Beijing 100049, China}

\author{Tian-Min Yan}
	\email{yantm@sari.ac.cn}
	\affiliation{Shanghai Advanced Research Institute, Chinese Academy of Sciences, Shanghai 201210, China}

\author{Y. H. Jiang}
	\email{jiangyh@sari.ac.cn}
	\affiliation{Shanghai Advanced Research Institute, Chinese Academy of Sciences, Shanghai 201210, China}
    \affiliation{University of Chinese Academy of Sciences, Beijing 100049, China}
	\affiliation{ShanghaiTech University, Shanghai 201210, China}

\begin{abstract}
	The study of quantum evolution on graphs for diversified topologies is beneficial to modeling various realistic systems. A systematic method, the dimerized decomposition, is proposed to analyze the dynamics on an arbitrary network. By introducing global "flows" among interlinked dimerized subsystems, each of which locally consists of an input and a output port, the method provides an intuitive picture that the local properties of the subsystem are separated from the global structure of the network. The pictorial interpretation of quantum evolution as multiple flows through the graph allows for the analysis of the complex network dynamics supplementary to the conventional spectral method.
\end{abstract}


\maketitle

\section{Introduction}

The quantum evolution on a network, which consists of multiple sites
and edges representing inter-site couplings, appeals increasing interests
for its wide applications ranging from quantum information \cite{farhi_quantum_1998}
and computation \cite{childs_universal_2009} to excitation transfer
\cite{valkunas_molecular_2013}. Typically, the quantity of interest
is the transport efficiency or the transfer time to specific site(s),
e.g., the maximized probability at the target site in the shortest
time for the spatial search algorithm \cite{childs_spatial_2004},
the enhanced efficiency of energy transfer assisted by coherence among
chromophores in photosynthetic complexes \cite{mohseni_environment-assisted_2008},
and the maximum fidelity to transmit a quantum state in a spin-network
from one point to another \cite{bose_quantum_2003}. In general, the
processes can be rephrased within the theoretical framework of continuous-time
quantum walk (CTQW) \cite{mulken_continuous-time_2011}, which outperforms
the classical counterpart by exploiting interference among different
paths in a graph. The experimental implementations of CTWQ are
proposed or achieved on various platforms including ultracold Rydberg
atoms \cite{cote_quantum_2006,mulken_survival_2007}, tight-binding
graphene lattice \cite{foulger_quantum_2014,bohm_microwave_2015},
and optical waveguide lattices \cite{perets_realization_2008,aspuru-guzik_photonic_2012,qiang_efficient_2016}.

Within the framework of CTQW, the techniques of dimensionality reduction
that project the complete space spanned by sites of the original system
to an equivalent one, or a subspace, have usually been applied, e.g.,
the invariant subspace methods using the Lanczos algorithm for systems
of proper symmetry \cite{novo_systematic_2015}, diagrammatic approach
by degenerate perturbation theory \cite{janmark_global_2014,meyer_connectivity_2015,wong_diagrammatic_2015}.
These methods considerably reduce the complexity and provide simplified
pictures analogous to well-known problems, e.g., the linear chain
decomposition that transforms a dendrimer to a line \cite{salimi_continuous-time_2010}
or linear chains \cite{koda_equivalence_2015}, and transport equivalent
quantum networks mapping onto classical resistor networks \cite{sarkar_equivalent_2016}.

In this work, a reduction scheme, the dimerized decomposition, is
introduced to simplify the analysis of quantum evolution on graphs.
The approach diverts our attention from the amplitudes on sites towards
\emph{flows}, the relations among sites, within the graph. Given an
$N$-site graph with $M$ coupling edges, the method serves to decompose
the graph into $M$ subsystems, each of which includes only two sites.
Within the subsystem, the dynamics are governed by the equation of
motion (EOM) similar to an ordinary Schr\"{o}dinger equation with
its local Hamiltonian containing the information of site energies,
local coupling, and explicit numbers of connectivities. The two sites
within the subsystem form a pair of ports, via which the local subsystem
is connected to other subsystems through auxiliary boundary terms,
interpreted as inter-subsystem ``flows''. Once the relations of
amplitudes among subsystems are set, the flows are determined. More
specifically, the relations yield a series of matching conditions
in the form of a linear system encoded by the global topologies, and
the flows are obtained by solving the linear system. The method provides
an intuitive picture that may simplify the design or optimization
of desired quantities, e.g., the efficiency of quantum transport.

The work is organized as follows: in Sec. \ref{sec:Theory}, we introduce
the dimerized decomposition and the EOM of the subsystem after the
decomposition. The validity of the method is shown starting with Schr\"{o}dinger
equation for the quantum evolution on a generic graph. In Sec. \ref{sec:Applications},
two examples using the decomposition are presented. The explicit expression
of the EOMs of subsystems are shown in Sec. \ref{sub:diamond_graph}
with the mathcing conditions given in \ref{sec:Appendix} for a diamond
graph, then in Sec. \ref{sub:trimer-efficiency} the transport efficiency
of a trimer system is analyzed using the method.

\section{Theory of dimerized decomposition\label{sec:Theory}}

Given a undirected graph $G=(V,E)$ consisting of the vertex set $V$
and edge set $E$, we start by decomposing the full system into subsystems
$\{\mathcal{S}\}$, each of which is associated with a local Hamiltonian
$\hat{H}^{(\mathcal{S})}$. The dimerized decomposition gains its
name from the scheme that each subsystem is constructed from the pair
of coupled sites, namely, the edge $\mathcal{S}=(i\sim j)\in E$.
In general, the subsystems are allowed to communicate with each other
and the mechanism can be realized via inter-subsystem flows. With
the above setup, it is shown that the original full Schr\"{o}dinger
equation for the quantum evolution on the graph can be casted into
a set of coupled EOMs of subsystems $\{\mathcal{S}\}$.

\begin{figure}
\includegraphics[scale=0.4]{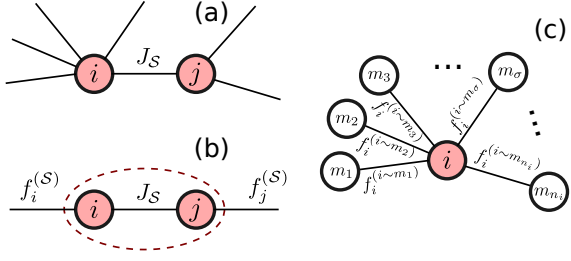}\caption{The scheme of dimerized decomposition in a graph. An arbitrary pair
of sites $i$ and $j$ that are coupled by edge of $J_{\mathcal{S}}$,
as shown by (a), can be viewed to form a local subsystem (b), which
is effectively isolated from the global network structure if all effects
of inter-subsystem communication are solely described by flows of
two ports, the auxillary functions $f_{i}^{(\mathcal{S})}$ and $f_{j}^{(\mathcal{S})}$.
Figure (c) shows how the EOM of the subsystem formed by any pair of
coupled sites $i$ and $m_{\sigma}$ is derived. Assuming an arbitrary
site $i$ in the graph is connected to $n_{i}$ neighboring sites,
for each connection between site $i$ and $m_{\sigma}$ there introduced
an auxiliary function $f_{i}^{(i\sim m_{\sigma})}$. \label{fig:eom-derivation}}
\end{figure}

As shown in Fig. \ref{fig:eom-derivation}(a), we consider the coupled
sites $i$ and $j$ with energies $\varepsilon_{i}$ and $\varepsilon_{j}$,
respectively. The coupling strength of the associated edge is $J_{\mathcal{S}}$.
Sites $i$ and $j$ together with edge $\mathcal{S}$ form the primitive
subsystem $\mathcal{S}$ as shown in Fig. \ref{fig:eom-derivation}(b).
Besides the intra-subsystem coupling $J_{\mathcal{S}}$, the two sites
may also be connected to sites outside $\mathcal{S}$. These multiple
connections except for $J_{\mathcal{S}}$ may be simplified by an
equivalent term, defined in our work as the time-dependent flow function
$f_{i}^{(\mathcal{S})}(t)$. As appeared in Fig. \ref{fig:eom-derivation}(b),
flows $f_{i}^{(\mathcal{S})}(t)$ and $f_{j}^{(\mathcal{S})}(t)$
are introduced for sites $i$ and $j$, respectively. In the following
we show the validity of such decomposition and derive the EOM within
the subsystem in terms of flows.

Without loss of generality, we start the decomposition from a single
arbitrary site $i$ as shown in Fig. \ref{fig:eom-derivation}. Let
$c_{i}(t)$ be the amplitude of site $i$ in the full system (i.e.,
before the decomposition). With the total number of connections of
site $i$ defined by connectivity $n_{i}$, all edges coupled to $i$
form a set $\mathcal{\mathscr{S}}_{i}=\{i\sim m_{\sigma}|\sigma=1,\cdots,n_{i}\}$.
From Schr\"{o}dinger equation, the EOM for site $i$ reads 
\begin{eqnarray}
\dot{c}_{i}(t) & = & -i\varepsilon_{i}c_{i}(t)-i\sum_{\sigma=1}^{n_{i}}J_{i\sim m_{\sigma}}c_{m_{\sigma}}(t),\label{eq:eom-original-full}
\end{eqnarray}
where $i\sim m_{\sigma}$ denotes site $i$ is connected to site $m_{\sigma}$.
For brevity, time variable $t$ (and variable $s$ in the Laplace
$s$-domain as will be introduced later) is henceforth dropped from
functions, unless noted otherwise. 

Let $c_{i}=\sum_{\mathcal{S}\in\mathscr{S}}c_{i}^{(\mathcal{S})}$
and substitute the sum into Eq. (\ref{eq:eom-original-full}), 
\begin{eqnarray}
\sum_{\mathcal{S}\in\mathscr{S}_{i}}\dot{c}_{i}^{(\mathcal{S})} & = & -i\varepsilon_{i}\sum_{\mathcal{S}\in\mathscr{S}_{i}}c_{i}^{(\mathcal{S})}-i\sum_{\sigma=1}^{n_{i}}J_{i\sim m_{\sigma}}\sum_{\mathcal{S}\in\mathscr{S}_{m_{\sigma}}}c_{m_{\sigma}}^{(\mathcal{S})}.\label{eq:eom-original-full-sum}
\end{eqnarray}
With the aim to separate the component $c_{i}^{(\mathcal{S})}=c_{i}^{(i\sim m_{1})}$
from the sum, we introduce an auxiliary function $f_{i}^{(i\sim m_{1})}$
and Eq. (\ref{eq:eom-original-full-sum}) is split as followings,
\begin{eqnarray}
\dot{c}_{i}^{(i\sim m_{1})} & = & -i\varepsilon_{i}c_{i}^{(i\sim m_{1})}-iJ_{i\sim m_{1}}\sum_{\mathcal{S}\in\mathscr{S}_{m_{1}}}c_{m_{1}}^{(\mathcal{S})}\nonumber \\
 &  & +f_{i}^{(i\sim m_{1})},\label{eq:eom-separation-1}\\
\sum_{\substack{\mathcal{S}\in\mathscr{S}_{i}\\
\mathcal{S}\neq(i\sim m_{1})
}
}\dot{c}_{i}^{(\mathcal{S})} & =-i & \varepsilon_{i}\sum_{\substack{\mathcal{S}\in\mathscr{S}_{i}\\
\mathcal{S}\neq(i\sim m_{1})
}
}c_{i}^{(\mathcal{S})}-i\sum_{\sigma=2}^{n_{i}}J_{i\sim m_{\sigma}}\sum_{\mathcal{S}\in\mathscr{S}_{\sigma}}c_{m_{\sigma}}^{(\mathcal{S})}\nonumber \\
 &  & -f_{i}^{(i\sim m_{1})}.\label{eq:eom-separation-2}
\end{eqnarray}
Only connection $i\sim m_{1}$ is contained in Eq. (\ref{eq:eom-separation-1}).
Subsequently, if we introduce another auxiliary function $f_{i}^{(i\sim m_{2})}$,
the EOM for exclusive connection of $i\sim m_{2}$ can also be separated
from Eq. (\ref{eq:eom-separation-2}) using the similar procedure.
Repeatedly, a series of equations for site $i$ of the whole set of
connections, $\mathscr{S}_{i}$, are derived,

\begin{eqnarray}
\dot{c}_{i}^{(i\sim m_{\sigma})} & = & -i\varepsilon_{i}c_{i}^{(i\sim m_{\sigma})}-iJ_{i\sim m_{\sigma}}\sum_{\mathcal{S}\in\mathscr{S}_{m_{\sigma}}}c_{m_{\sigma}}^{(\mathcal{S})}\nonumber \\
 &  & +f_{i}^{(i\sim m_{\sigma})},\label{eq:eom-decomposition-general}
\end{eqnarray}
for $\sigma=1,\cdots,n_{i}$. Each connection $i\sim m_{\sigma}$
is associated to a dimerized subsystem. The auxiliary functions $f_{i}^{(i\sim m_{\sigma})}$
for all edges need satisfy the requirement
\begin{eqnarray}
\sum_{\sigma=1}^{n_{i}}f_{i}^{(i\sim m_{\sigma})} & = & 0,\label{eq:junction-rule}
\end{eqnarray}
similar to Kirchhoff's junction rule for DC circuits that the net
flow (the sum over all flows for site $i$) at a junction is zero.

In Eq. (\ref{eq:eom-decomposition-general}), the EOMs of $n_{i}$
local subsystems are exact and no extra assumption is introduced.
However, we are still free to choose the form of $c_{i}^{(\mathcal{S})}$
in the sum in Eq. (\ref{eq:eom-decomposition-general}), and the auxiliary
functions $f_{i}^{(i\sim m_{\sigma})}$ should be reversely influenced
by the choice. Since amplitudes $c_{i}^{(\mathcal{S})}$ in subsystems
are desired to reflect the actual amplitude $c_{i}$ in the full system,
and we also wish to treat the sum with further simplicity, it is natural
to impose the assumption 
\begin{equation}
c_{i}^{(i\sim m_{1})}=c_{i}^{(i\sim m_{2})}=\cdots=c_{i}^{(i\sim m_{n_{i}})}.\label{eq:matching-condition}
\end{equation}
Thus, the relation of amplitudes between the full system and the subsystems
is simply $c_{i}=\sum_{\mathcal{S}\in\mathscr{S}_{i}}c_{i}^{(\mathcal{S})}=n_{i}c_{i}^{(i\sim m_{\sigma})}$
for any of the $\sigma$th subsystem. Thereby substituting into Eq.
(\ref{eq:eom-decomposition-general}) results in
\begin{eqnarray}
\dot{c}_{i}^{(i\sim m_{\sigma})} & = & -i\varepsilon_{i}c_{i}^{(i\sim m_{\sigma})}-in_{m_{\sigma}}J_{i\sim m_{\sigma}}c_{m_{\sigma}}^{(i\sim m_{\sigma})}\nonumber \\
 &  & +f_{i}^{(i\sim m_{\sigma})}.\label{eq:eom-decomposition}
\end{eqnarray}
Besides the outcome of simplified EOMs, the equal distribution of
amplitudes $c_{i}^{(\mathcal{S})}$ in $n_{i}$ subsystems establishes
the matching conditions among subsystems, which is a critical step
to find $f_{i}^{(\mathcal{S})}$. The $n_{i}-1$ equations from matching
conditions in Eq. (\ref{eq:matching-condition}), together with the
junction rule of Eq. (\ref{eq:junction-rule}), can uniquely determine
the $n_{i}$ functions $f_{i}^{(\mathcal{S})}$.

The above EOMs are derived from the perspective of a single site $i$
for an arbitrary graph. Each equation describes the evolution of a
coupling edge $\mathcal{S}=(i\sim m_{\sigma})\in\mathscr{S}_{i}$.
Extending the single site to all sites $i\in V$, the above derivation
may reversely be viewed from the perspective of edges instead of sites.
If sites $i$ and $j$ are connected, it is always possible to select
a pair of EOMs for $c_{i}^{(i\sim j)}$ and $c_{j}^{(j\sim i)}$ from
Eq. (\ref{eq:eom-decomposition}),
\begin{eqnarray*}
\dot{c}_{i}^{(i\sim j)} & = & -i\varepsilon_{i}c_{i}^{(i\sim j)}-in_{j}J_{i\sim j}c_{j}^{(i\sim j)}+f_{i}^{(i\sim j)},\\
\dot{c}_{j}^{(j\sim i)} & = & -i\varepsilon_{j}c_{j}^{(j\sim i)}-in_{i}J_{j\sim i}c_{i}^{(j\sim i)}+f_{j}^{(j\sim i)},
\end{eqnarray*}
which describes the dynamics within subsystem $\mathcal{S}=(i\sim j)=(j\sim i)$.
A more compact matrix form is 
\begin{equation}
\overset{.}{\boldsymbol{c}}^{(\mathcal{S})}=-i\boldsymbol{H}^{(\mathcal{S})}\boldsymbol{c}^{(\mathcal{S})}+\boldsymbol{f}^{(\mathcal{S})},\label{eq:eom-decomposition-matrix-form}
\end{equation}
with $\boldsymbol{c}^{(\mathcal{S})}=\left(c_{i}^{(\mathcal{S})},c_{j}^{(\mathcal{S})}\right)^{\mathrm{T}}$,
the local Hamiltonian
\begin{equation}
\boldsymbol{H}^{(\mathcal{S})}=\begin{pmatrix}\varepsilon_{i} & n_{j}J_{\mathcal{S}}\\
n_{i}J_{\mathcal{S}} & \varepsilon_{j}
\end{pmatrix},\label{eq:local-hamiltonian}
\end{equation}
and the boundary term $\boldsymbol{f}^{(\mathcal{S})}=\left(f_{i}^{(\mathcal{S})},f_{j}^{(\mathcal{S})}\right)^{\mathrm{T}}$
accounting for flows via sites in subsystem $\mathcal{S}$. Although
$\boldsymbol{H}^{(\mathcal{S})}$ is unsymmetric when $n_{i}\neq n_{j}$,
the hermiticity can still be conserved with a new set of $\boldsymbol{c}^{(\mathcal{S})}$
by proper linear transformation. When $\boldsymbol{f}^{(\mathcal{S})}=\left(0,0\right)^{\mathrm{T}}$,
Eq. (\ref{eq:eom-decomposition-matrix-form}) is essentially the Schr\"{o}dinger
equation for the two-level system with the off-diagonal couplings
modified by connectivities.

Given the initial condition $\boldsymbol{c}^{(\mathcal{S})}(0)$,
the formal solution of Eq. (\ref{eq:eom-decomposition-matrix-form})
reads

\begin{equation}
\boldsymbol{c}^{(\mathcal{S})}(t)=\int_{0}^{t}\boldsymbol{U}^{(\mathcal{S})}(t-\tau)\boldsymbol{f}^{(\mathcal{S})}(\tau)d\tau+\boldsymbol{U}^{(\mathcal{S})}(t)\boldsymbol{c}^{(\mathcal{S})}(0),\label{eq:ct-of-local-eom}
\end{equation}
where $\boldsymbol{U}^{(\mathcal{S})}(t)=e^{-i\boldsymbol{H}^{(\mathcal{S})}t}$
is the local time-evolution operator of subsystem $\mathcal{S}$. 

Our aim is to find $\boldsymbol{f}^{(\mathcal{S})}$ which eventually
determines amplitude $\boldsymbol{c}^{(\mathcal{S})}$. Since Eq.
(\ref{eq:ct-of-local-eom}) takes the form of a Volterra integral
that can be easily analyzed after the Laplace transform,
\begin{equation}
\tilde{\boldsymbol{c}}^{(\mathcal{S})}(s)=\boldsymbol{\tilde{U}}^{(\mathcal{S})}(s)[\boldsymbol{\tilde{f}}^{(\mathcal{S})}(s)+\boldsymbol{c}^{(\mathcal{S})}(0)],\label{eq:cs-of-local-eom}
\end{equation}
the calculation of $\boldsymbol{f}^{(\mathcal{S})}$ is actually conducted
in the $s$-domain. Here, variables in the $s$-domain, as appeared
in Eq. (\ref{eq:cs-of-local-eom}), are indicated by the tilde. The
local time evolution operator $\tilde{\boldsymbol{U}}^{(\mathcal{S})}(s)$
in the $s$-domain corresponding to $\boldsymbol{H}^{(\mathcal{S})}$
in Eq. (\ref{eq:local-hamiltonian}) reads
\begin{equation}
\tilde{\boldsymbol{U}}^{(\mathcal{S})}(s)=\frac{1}{\Omega^{2}+\bar{s}^{2}}\begin{pmatrix}\bar{s}-i\Delta_{ij} & -in_{j}J_{\mathcal{S}}\\
-in_{i}J_{\mathcal{S}} & \bar{s}+i\Delta_{ij}
\end{pmatrix}\label{eq:Us}
\end{equation}
with $\bar{s}=s+i\bar{\varepsilon}$ , $\Delta_{ij}=(\varepsilon_{i}-\varepsilon_{j})/2$,
$\bar{\varepsilon}=(\varepsilon_{i}+\varepsilon_{j})/2$ and $\Omega=\sqrt{\Delta_{ij}^{2}+n_{i}n_{j}J_{\mathcal{S}}^{2}}$. 

In the $s$-domain, the flow function $\boldsymbol{f}^{(\mathcal{S})}$
can be uniquely determined by solving the linear system generated
from both the junction rule Eq. (\ref{eq:junction-rule}) and matching
conditions Eq. (\ref{eq:matching-condition}). In Eq. (\ref{eq:junction-rule}),
the junction rules are directly expressed as equalities among $f_{i}^{(\mathcal{S})}$.
The explicit form of matching condition Eq. (\ref{eq:matching-condition}),
however, assuming site $i$ is shared by both subsystems $\mathcal{S}$
and $\mathcal{T}$, is given by
\begin{eqnarray}
 &  & \tilde{u}_{\sigma,1}^{(\mathcal{S})}[\tilde{f}_{i_{1}^{(\mathcal{S})}}^{(\mathcal{S})}+c_{i_{1}^{(\mathcal{S})}}(0)]+\tilde{u}_{\sigma,2}^{(\mathcal{S})}[\tilde{f}_{i_{2}^{(\mathcal{S})}}^{(\mathcal{S})}+c_{i_{2}^{(\mathcal{S})}}(0)]\nonumber \\
 & = & \tilde{u}_{\tau,1}^{(\mathcal{T})}[\tilde{f}_{i_{1}^{(\mathcal{T})}}^{(\mathcal{T})}+c_{i_{1}^{(\mathcal{T})}}(0)]+\tilde{u}_{\tau,2}^{(\mathcal{T})}[\tilde{f}_{i_{2}^{(\mathcal{T})}}^{(\mathcal{T})}+c_{i_{2}^{(\mathcal{T})}}(0)],\label{eq:matching-condition-explicit}
\end{eqnarray}
where $\tilde{u}_{\sigma(\tau),1(2)}^{(\mathcal{S})}$ is the matrix
element of $\tilde{\boldsymbol{U}}^{(\mathcal{S})}$ in Eq. (\ref{eq:Us}),
and $c_{i}(0)$ is the initial amplitude on site $i$. Indices $\sigma,\tau=1,2$
label the intra-dimer sites, and $i_{\alpha}^{(\mathcal{S})}$ with
$\alpha=1,2$ is the actual site index for the $\alpha$th site in
subsystem $\mathcal{S}$. Note that within the set $\{i_{1}^{(\mathcal{S})},i_{2}^{(\mathcal{S})},i_{1}^{(\mathcal{T})},i_{2}^{(\mathcal{T})}\}$,
two indices must be the same as specified by the matching condition
for $\mathcal{S}\neq\mathcal{T}$. Given the unknown flow functions
arranged by $\tilde{\boldsymbol{f}}=\left(\tilde{\boldsymbol{f}}^{(a)},\tilde{\boldsymbol{f}}^{(b)},\cdots,\tilde{\boldsymbol{f}}^{(N)}\right)^{\mathrm{T}}$
with the local flow for subsystem $\mathcal{S}$, $\tilde{\boldsymbol{f}}^{(\mathcal{S})}=\left(\tilde{f}_{i_{1}^{(\mathcal{S})}}^{(\mathcal{S})},\tilde{f}_{i_{2}^{(\mathcal{S})}}^{(\mathcal{S})}\right)^{\mathrm{T}}$,
$\tilde{\boldsymbol{f}}$ can be found by solving the linear system
$\tilde{\boldsymbol{M}}\tilde{\boldsymbol{f}}=\tilde{\boldsymbol{b}}$,
where $\tilde{\boldsymbol{M}}$ is the matrix constructed from the
matching condition and junction rule. Given an $N$-site graph with
$M$ coupling edges, $\tilde{\boldsymbol{M}}$ is a $2M\times2M$
matrix accounting for $N$ equations from junction rules and the rest
$2M-N$ equations from matching conditions. The global topology of
the graph is encoded in $\tilde{\boldsymbol{M}}$, whose matrix elements
are also embedded with the local properties of subsystems. The array
$\tilde{\boldsymbol{b}}$ is an array formed by all non-$\tilde{f}_{i}^{(\mathcal{S})}$
terms in Eq. (\ref{eq:matching-condition-explicit}) related to initial
conditions $c_{i}(0)$. In the following, the method will be presented
in detail with examples.

We note that, in regard to the computational complexity when solving
the differential equations, admittedly, our method is not advantageous.
Given a homogeneous system of the EOM $i\dot{\boldsymbol{c}}=\boldsymbol{H}\boldsymbol{c}$
with the $N$-site hamiltonian $\boldsymbol{H}$, the typical evaluation
of the wavefunctions by $\boldsymbol{c}(t)=\mathcal{T}e^{-i\int^{t}dt'\boldsymbol{H}(t')}\boldsymbol{c}(0)$
requires one to find the time evolution operator $e^{-i\boldsymbol{H}\Delta t}$,
equivalent to the spectral decomposition $\boldsymbol{U}^{\text{T}}e^{-i\boldsymbol{D}\Delta t}\boldsymbol{U}$.
In our method, the diagonalization of $\boldsymbol{H}$ is not required,
since all time evolution operator within the two-level subsystem has
a fixed-format closed form solution. Instead, solving the original
Schr\"{o}dinger equation is recast as treating a series of coupled
inhomogeneous two-dimensional matrix equations. The most computationally
demanding part is to find the inhomogeneous term $\boldsymbol{f}_{i}^{(\mathcal{S})}$
from $\sum_{\{i|\deg(v_{i})>1\}}\deg(v_{i})$ matching conditions.
Usually, the complexity of solving the linear system $\tilde{\boldsymbol{M}}\tilde{\boldsymbol{f}}=\tilde{\boldsymbol{b}}$
is even higher than the exact diagonalization of the original hamiltonian,
though the matching matrix is usually sparse because, as suggested
by Eq. (\ref{eq:matching-condition-explicit}), each row of $\tilde{\boldsymbol{M}}$
has at most four non-zero elements.

\section{Applications\label{sec:Applications}}

\subsection{Flow patterns in diamond graph\label{sub:diamond_graph}}

\begin{figure}
\includegraphics[scale=0.35]{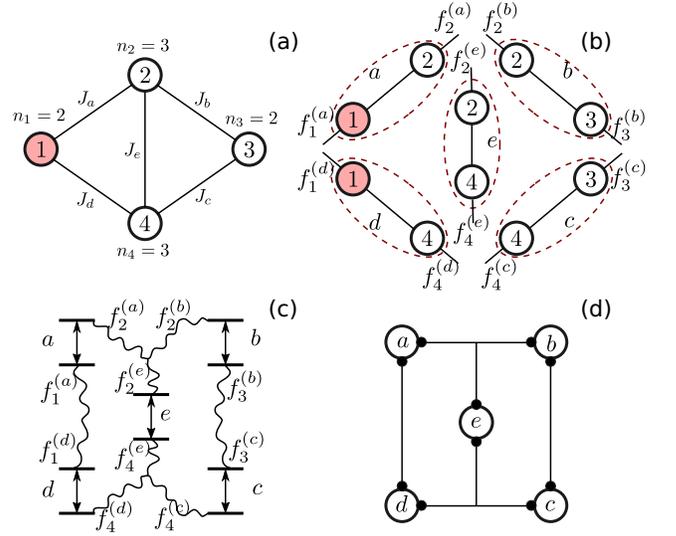}

\caption{(a) A four-site diamond graph. The initial state is on site $|1\rangle$
labeled in red. (b) The decomposition of the graph according to the
coupling edges into subsystems as indicated by dashed circles. The
flow functions $f_{i}^{(\mathcal{S})}$ are labeled by the double
ends of each subsystem. (c) Alternatively, each subspace as a two-level
system represents a qubit subject to perturbing flows on the two internal
states. (d) The dimerized decomposition is equivalent to swapping
the roles of sites and edges in panel (a), and now we focus on the
central role of edges in the original graph. Each site (circle) is
a dimerized subspace, whose internal sites are perturbed by auxiliary
flows represented by a pair of black dots. \label{fig:diamond_graph}}
\end{figure}

In order to show the procedure of the decomposition and obtain the
flow patterns, the method is applied to a diamond graph (2-fan including
four sites) as shown in Fig. \ref{fig:diamond_graph}(a). The Hamiltonian
is $\hat{H}=\sum_{i}^{4}\varepsilon_{i}|i\rangle\langle i|+J_{a}|1\rangle\langle2|+J_{b}|2\rangle\langle3|+J_{c}|3\rangle\langle4|+J_{d}|4\rangle\langle1|+J_{e}|2\rangle\langle4|+\mathrm{c}.\mathrm{c}.$
and the amplitude on site $i$ of the full system is $c_{i}(t)$.
A diagrammatic representation of the decomposition as shown in Fig.
\ref{fig:diamond_graph}(b) allows for the direct translation of Eq.
(\ref{eq:eom-decomposition-matrix-form}) for local EOMs of subsystems
$\mathcal{S}\in\{a,b,c,d,e\}$ as followings,

\begin{eqnarray}
\left(\begin{array}{c}
\dot{c}_{1}^{(a)}\\
\dot{c}_{2}^{(a)}
\end{array}\right) & = & -i\left(\begin{array}{cc}
\varepsilon_{1} & 3J_{a}\\
2J_{a} & \varepsilon_{2}
\end{array}\right)\left(\begin{array}{c}
c_{1}^{(a)}\\
c_{2}^{(a)}
\end{array}\right)+\left(\begin{array}{c}
f_{1}^{(a)}\\
f_{2}^{(a)}
\end{array}\right),\nonumber \\
\left(\begin{array}{c}
\dot{c}_{2}^{(b)}\\
\dot{c}_{3}^{(b)}
\end{array}\right) & = & -i\left(\begin{array}{cc}
\varepsilon_{2} & 2J_{b}\\
3J_{b} & \varepsilon_{3}
\end{array}\right)\left(\begin{array}{c}
c_{2}^{(b)}\\
c_{3}^{(b)}
\end{array}\right)+\left(\begin{array}{c}
f_{2}^{(b)}\\
f_{3}^{(b)}
\end{array}\right),\nonumber \\
\left(\begin{array}{c}
\dot{c}_{3}^{(c)}\\
\dot{c}_{4}^{(c)}
\end{array}\right) & = & -i\left(\begin{array}{cc}
\varepsilon_{3} & 3J_{c}\\
2J_{c} & \varepsilon_{4}
\end{array}\right)\left(\begin{array}{c}
c_{3}^{(c)}\\
c_{4}^{(c)}
\end{array}\right)+\left(\begin{array}{c}
f_{3}^{(c)}\\
f_{4}^{(c)}
\end{array}\right),\nonumber \\
\left(\begin{array}{c}
\dot{c}_{4}^{(d)}\\
\dot{c}_{1}^{(d)}
\end{array}\right) & = & -i\left(\begin{array}{cc}
\varepsilon_{4} & 2J_{d}\\
3J_{d} & \varepsilon_{1}
\end{array}\right)\left(\begin{array}{c}
c_{4}^{(d)}\\
c_{1}^{(d)}
\end{array}\right)+\left(\begin{array}{c}
f_{4}^{(d)}\\
f_{1}^{(d)}
\end{array}\right),\nonumber \\
\left(\begin{array}{c}
\dot{c}_{2}^{(e)}\\
\dot{c}_{4}^{(e)}
\end{array}\right) & = & -i\left(\begin{array}{cc}
\varepsilon_{2} & 3J_{e}\\
3J_{e} & \varepsilon_{4}
\end{array}\right)\left(\begin{array}{c}
c_{2}^{(e)}\\
c_{4}^{(e)}
\end{array}\right)+\left(\begin{array}{c}
f_{2}^{(e)}\\
f_{4}^{(e)}
\end{array}\right),\label{eq:eom-of-diamond-graph}
\end{eqnarray}
where $c_{i}^{(\mathcal{S})}$ is the amplitude in subsystem. According
to matching conditions, we have relations of amplitudes between subsystems
and the full system, $c_{i}^{(\mathcal{S})}=c_{i}/2$ for $i=1,3$
and $c_{i}^{(\mathcal{S})}=c_{i}/3$ for $i=2,4$.

Next we show how flow functions in Eq. (\ref{eq:eom-of-diamond-graph})
of the form $\tilde{\boldsymbol{f}}=\left(\tilde{\boldsymbol{f}}^{(a)},\tilde{\boldsymbol{f}}^{(b)},\tilde{\boldsymbol{f}}^{(c)},\tilde{\boldsymbol{f}}^{(d)},\tilde{\boldsymbol{f}}^{(e)}\right)^{\mathrm{T}}=\left(\tilde{f}_{1}^{(a)},\tilde{f}_{2}^{(a)},\tilde{f}_{2}^{(b)},\tilde{f}_{3}^{(b)},\tilde{f}_{3}^{(c)},\tilde{f}_{4}^{(c)},\tilde{f}_{4}^{(d)},\tilde{f}_{1}^{(d)},\tilde{f}_{2}^{(e)},\tilde{f}_{4}^{(e)}\right)^{\mathrm{T}}$
are determined. As in Eq. (\ref{eq:matching-condition}), the six
restricting equalities, $c_{1}^{(a)}=c_{1}^{(d)}$, $c_{2}^{(a)}=c_{2}^{(b)}=c_{2}^{(e)}$,
$c_{3}^{(b)}=c_{3}^{(c)}$ and $c_{4}^{(c)}=c_{4}^{(d)}=c_{4}^{(e)}$,
are imposed by matching conditions. Together with the four equations
from the junction rules, $f_{1}^{(a)}+f_{1}^{(d)}=0$, $f_{2}^{(a)}+f_{2}^{(b)}+f_{2}^{(d)}=0$,
$f_{3}^{(b)}+f_{3}^{(c)}=0$ and $f_{4}^{(c)}+f_{4}^{(d)}+f_{4}^{(e)}=0$,
we construct the matrix $\tilde{\boldsymbol{M}}$ (see Appendix for
the explicit form) from which the ten flow functions can be determined
by solving $\tilde{\boldsymbol{f}}=\tilde{\boldsymbol{M}}^{-1}\tilde{\boldsymbol{b}}$. 

Taking the simplest case with all edges of identical coupling strength
$J$ for instance, solving the determinant equation $|\tilde{\boldsymbol{M}}(s)|=0$
shows the $k$th zero is given by $s_{k}^{(p)}\in iJ\{-\frac{1}{2}(\sqrt{17}+1),0,1,\frac{1}{2}(\sqrt{17}-1)\}$.
Since $\tilde{\boldsymbol{f}}=\tilde{\boldsymbol{M}}^{-1}\tilde{\boldsymbol{b}}$,
$s_{k}^{(p)}$ is a pole of $\tilde{\boldsymbol{f}}$ in the complex
$s$-plane. Clearly, the relation between the $k$th pole with the
eigenvalue $\lambda_{k}$ obtained using spectral method is given
by $\lambda_{k}=is_{k}^{(p)}$.

Given a subsystem $\mathcal{S}$, from Eq. (\ref{eq:cs-of-local-eom})
it is clearly seen that the amplitude $\tilde{\boldsymbol{c}}^{(\mathcal{S})}$,
as a vector spanned in the basis of the local subsystem, can be obtained
from the driving source $\tilde{\boldsymbol{f}}^{(\mathcal{S})}$
in Eq. (\ref{eq:cs-of-local-eom}) followed by the vector operation
of the local time evolution $\tilde{\boldsymbol{U}}^{(S)}$. Usually,
only poles of $\tilde{\boldsymbol{c}}^{(\mathcal{S})}$ in the $s$-plane,
which are equivalent to the poles of $\tilde{\boldsymbol{f}}^{(\mathcal{S})}$,
contributes to $\boldsymbol{c}^{(\mathcal{S})}$ in the time domain.
The poles are typically directly derived from the zeros of the characteristic
polynomial $|\tilde{\boldsymbol{M}}(s)|$. But one should note that
the value that renders any matrix element of $\tilde{\boldsymbol{M}}$
singular and coincides with any eigenvalue of $\tilde{\boldsymbol{U}}^{(S)}$
is also a pole. For each pole $s_{k}^{(p)}$, the real and imaginary
parts represent the decay rate and oscillation frequency, respectively.
It corresponds to an eigenmode of $\lambda_{k}$ or a path in the
spectral method. The amplitude $\boldsymbol{c}^{(\mathcal{S})}$ is
the superposition of components over all these modes. Each as a mode
should present a distribution chart of $\text{Res}_{s_{k}^{(p)}}\tilde{f}_{i}$
as illustrated in Fig. \ref{fig:diamond_graph}(b). Here, $\text{Res}_{s_{k}^{(p)}}\tilde{f}_{i}$
is the residue of $\tilde{f}_{i}$ for the $k$th pole $s_{k}^{(p)}=-i\lambda_{k}$.

Although the system can be analyzed with the spectral method as well,
the added values of the method is that it provides a different perspective
to view the quantum evolution based on edges of a graph. Conventionally,
sites or vertices are considered the primitive and one typically focuses
on the evolution of components of sites. Here, however, edges are
viewed as taking the central role, as shown in Fig. \ref{fig:diamond_graph}(d).
The subspace, consisting of a pair of sites and the coupling edge,
is a two-level qubit that is the smallest nontrivial local system
{[}Fig. \ref{fig:diamond_graph}(c){]} with tremendously wide applications.
Unlike a usual isolated two-level system, however, states $i$ and
$j$ of $\mathcal{S}$ are allowed to be perturbed by auxiliary functions
$f_{i}^{(\mathcal{S})}(t)$ and $f_{j}^{(\mathcal{S})}(t)$, respectively.
As depicted by the matching condition, $f_{i}^{(\mathcal{S})}$ is
not arbitrarily but necessarily introduced to tune the amplitudes
in the local qubit to be consistent with the ones in the original
network. Since $f_{i}^{(\mathcal{S})}$ can be uniquely determined
once the topology and parameters of the network is given, it is characteristic
of the dynamics on the network.

An intuitive way to understand the role of $f_{i}^{(\mathcal{S})}$
is to visualize the distribution over all sites and edges on the graph.
In the time domain, $f_{i}^{(\mathcal{S})}(t)$ being a time-dependent
continuous function is difficult to present for a static image of
the network, therefore it is helpful to switch to the $s$-domain
and seek for an appropriate representation. In fact, as the amplitude
of site $i$ is given by $c_{i}(t)=\sum_{k}C_{i,k}e^{-i\lambda_{k}t}$
for $\lambda_{k}=is_{k}^{(p)}$ with $k$ over all eigenmodes, only
a finite numbers of $s_{k}^{(p)}$ contribute to the wave function
after transforming back to the time domain. However, since $f_{i}^{(\mathcal{S})}(s)$
at a pole is singular, the distribution of $\chi_{i,k}^{(\mathcal{S})}\equiv\mathrm{Res}_{s\rightarrow s_{k}^{(p)}}f_{i}^{(\mathcal{S})}(s)$
is instead shown on the graph.

For the $k$th eigenvalue $\lambda_{k}=is_{k}^{(p)}$, defining the
vector $\boldsymbol{C}_{k}^{(\mathcal{S})}\equiv(C_{i,k},C_{j,k})^{\mathrm{T}}$
for the pair of sites within $\mathcal{S}=i\sim j$, it is shown from
Eq. (\ref{eq:cs-of-local-eom}) that 
\begin{eqnarray}
\boldsymbol{C}_{k}^{(\mathcal{S})} & = & \mathrm{Res}_{s\rightarrow s_{k}^{(p)}}\tilde{\boldsymbol{c}}^{(\mathcal{S})}(s)=\tilde{\boldsymbol{U}}^{(\mathcal{S})}(s_{k}^{(p)})\boldsymbol{\chi}_{k}^{(\mathcal{S})},\label{eq:coefficient-vector}
\end{eqnarray}
where $\boldsymbol{\chi}_{k}^{(\mathcal{S})}\equiv\mathrm{Res}_{s\rightarrow s_{k}^{(p)}}\tilde{\boldsymbol{f}}^{(\mathcal{S})}(s)$.
In other words, coefficient $\boldsymbol{C}_{k}^{(\mathcal{S})}$
is the response of the local evolution operator $\tilde{\boldsymbol{U}}^{(\mathcal{S})}(s_{k}^{(p)})$
to the whole network determined perturbing source $\boldsymbol{\chi}_{k}^{(\mathcal{S})}$.
Eq. (\ref{eq:coefficient-vector}) effectively separates the influence
of the global network outside the local system from the one within
the local system. Especially when $\tilde{\boldsymbol{U}}^{(\mathcal{S})}(s_{k}^{(p)})$
of the studied subsystem remains untouched, it allows one to trace
how $\boldsymbol{C}_{k}^{(\mathcal{S})}$ is affected sololy by the
global network determined $\boldsymbol{\chi}_{k}^{(\mathcal{S})}$.

\begin{figure*}
\includegraphics[scale=0.315]{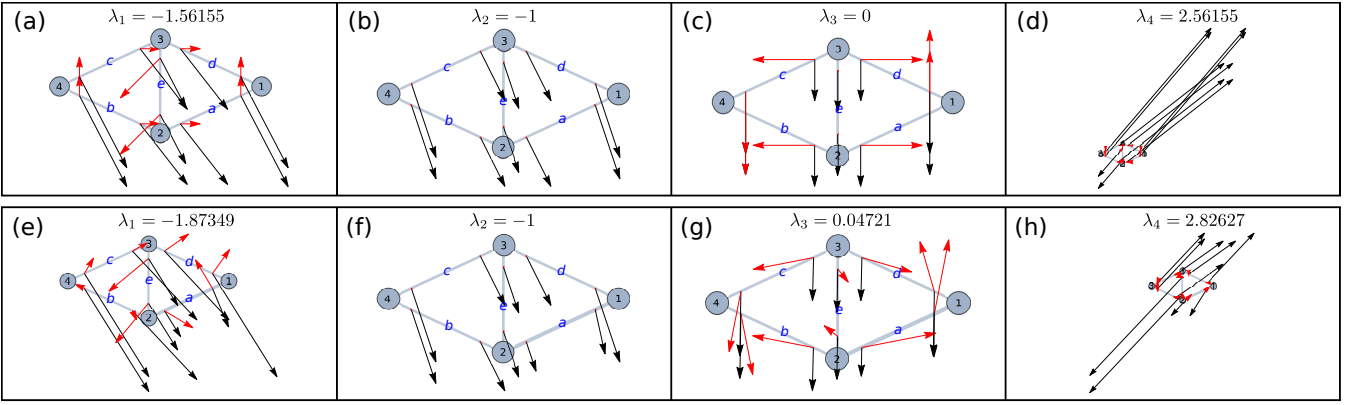}
\caption{Distribution of vectors that separate the local response from the
global-network induced perturbation in the diamond graph of Fig. \ref{fig:diamond_graph}.
Panels (a)-(d) ordered by eigenvalues $\lambda_{k}$ from lower to
higher values show the distribution when coupling strengths are all
$J_{\mathcal{S}}=1$. On all sites $\{i\}$ and edges $\{\mathcal{S}\}$,
vectors $\boldsymbol{u}_{i,k}^{(\mathcal{S})}$ (black arrow) and
$\boldsymbol{\chi}_{i,k}^{(\mathcal{S})}$ (red arrow) represent contributions
from local evolution operator and perturbing flow, respectively. The
vectors are defined as followings: If site $i$ is the first site
within $\mathcal{S}$, we define $\boldsymbol{u}_{i,k}^{(\mathcal{S})}=[\tilde{u}_{1,1}^{(\mathcal{S})}(s_{k}^{(p)}),\tilde{u}_{1,2}^{(\mathcal{S})}(s_{k}^{(p)})]$.
Otherwise, if site $i$ is treated as the second site in $\mathcal{S}$,
$\boldsymbol{u}_{i,k}^{(\mathcal{S})}=[\tilde{u}_{2,2}^{(\mathcal{S})}(s_{k}^{(p)}),\tilde{u}_{2,1}^{(\mathcal{S})}(s_{k}^{(p)})]$.
Similarly, the vector of the perturbing source at site $i$ is defined
by $\boldsymbol{\chi}_{i,k}^{(\mathcal{S})}=[\chi_{i,k}^{(\mathcal{S})},\chi_{j,k}^{(\mathcal{S})}]$
if $\mathcal{S}$ has two sites $i$ and $j$. The above definitions
always render the studied site as the first element that plays the
central role. Vectors $\boldsymbol{u}_{i,k}^{(\mathcal{S})}$ and
$\boldsymbol{\chi}_{i,k}^{(\mathcal{S})}$ are labeled on each edge
$\mathcal{S}$ around site $i$. Moreover, as Eq. (\ref{eq:coefficient-vector})
suggests, the coefficient of the wave function can be read directly
from the inner product of the pair of $\boldsymbol{u}_{i,k}^{(\mathcal{S})}$
and $\boldsymbol{\chi}_{i,k}^{(\mathcal{S})}$ (projection from black
arrow to the red arrow). Panels (e)-(g) show the vector distributions
when $J_{a}=1.5$ while all other edges remain $J_{\mathcal{S}}=1$
for $\mathcal{S}\protect\neq a$.\label{fig:arrow-patterns-diamond}}
\end{figure*}

The visualization of both contributions from $\tilde{\boldsymbol{U}}^{(\mathcal{S})}(s_{k}^{(p)})$
and $\boldsymbol{\chi}_{k}^{(\mathcal{S})}$ over all sites and edges
are shown in Fig. \ref{fig:arrow-patterns-diamond}. With full definitions
listed in the caption of the figure, the corresponding vectors are
redefined by $\boldsymbol{u}_{i,k}^{(\mathcal{S})}$ and $\boldsymbol{\chi}_{i,k}^{(\mathcal{S})}$
for site $i$, edge $\mathcal{S}$ and eigenmode $k$. The vectors
show the separation of the internal influence of the local subspace
from the external one outside the subspace. The vectors for eigenmode
$\lambda_{k}$ can be easily compared among all sites on different
edges. The magnitude of vector $\boldsymbol{\chi}_{i,k}^{(\mathcal{S})}$
represents the influence from the global network imposing on the subspace.
As for the directions of vectors $\boldsymbol{u}_{i,k}^{(\mathcal{S})}$
and $\boldsymbol{\chi}_{i,k}^{(\mathcal{S})}$, if subspace $\mathcal{S}$
includes sites $i$ and $j$, the larger the horizontal component
of a vector, the larger the contribution from the studied site $i$;
while the larger the component in the vertical direction, the larger
the contribution from the other site $j$. The direction of vectors
may help define the phase within the subsystem to study the change
with parameters. Due to the junction rule, $\sum_{\mathcal{S}}\boldsymbol{\chi}_{i,k}^{(\mathcal{S})}=0$,
for site $i$ the sum over horizontal components of $\boldsymbol{\chi}_{i,k}^{(\mathcal{S})}$
for different $\mathcal{S}$ is zero.

The separation of the contributions may be examined from two perspectives.
On one hand, when all parameters (e.g., site energies and coupling
strengths) are identical, the distribution of vectors characterizes
the influence of global topology on the quantum evolution of the nearest
neighboring environment. When the network structure is fixed, on the
other hand, if some local property is altered, the distribution informs
how the change of parameters perturbs local subsystems. Here, we present
the analysis following the guideline: Fig. \ref{fig:arrow-patterns-diamond}(a)-(d)
show the distribution of vectors for the diamond graphs with all parameters
identical, while (e)-(h) are results when the graph is perturbed by
changing $J_{a}=1.5$.

Vector $\boldsymbol{u}_{i,k}^{(\mathcal{S})}$ reveals local properties
within the subspace. Especially, as shown in Fig. \ref{fig:arrow-patterns-diamond}(a)-(d),
when $J_{\mathcal{S}}=1$ for all edges and the whole network is symmetric
along sites 1-2-3 and 1-4-3, vectors $\boldsymbol{u}_{2,k}^{(b)}$,
$\boldsymbol{u}_{3,k}^{(c)}$, $\boldsymbol{u}_{2,k}^{(a)}$ and $\boldsymbol{u}_{3,k}^{(d)}$
are identical due to the same local properties within subspaces $\mathcal{S}=a,b,c$
and $d$. Similarly, $\boldsymbol{u}_{2,k}^{(e)}$ and $\boldsymbol{u}_{3,k}^{(e)}$
are also the same in subspace $e$. But the latter two vectors differ
from the former because of the difference of connectivities, reflecting
properties of individual local subspaces.

The local properties of subspaces, including site energies, connectivities,
and the coupling strength, determine $\boldsymbol{u}_{i,k}^{(\mathcal{S})}$.
Though the coupling strength in $\mathcal{S}=a$ is changed in Fig.
\ref{fig:arrow-patterns-diamond}(e)-(h), the vectors outside $\mathcal{S}$,
like $\boldsymbol{u}_{2,k}^{(b)}$, $\boldsymbol{u}_{3,k}^{(c)}$
and $\boldsymbol{u}_{3,k}^{(d)}$, $\boldsymbol{u}_{4,k}^{(b)}$ and
$\boldsymbol{u}_{4,k}^{(c)}$, $\boldsymbol{u}_{2,k}^{(e)}$ and $\boldsymbol{u}_{3,k}^{(e)}$
are still identical because of the same local environments. The vectors
are only slightly changed when $J_{a}=1\rightarrow J_{a}=1.5$ by
different eigenvalues. While vectors within $a$, $\boldsymbol{u}_{1,k}^{(a)}$
and $\boldsymbol{u}_{2,k}^{(a)}$, change dramatically.

Vector $\boldsymbol{\chi}_{i,k}^{(\mathcal{S})}$ denotes the global
influence of the whole network on the local subspace. For some eigenmode,
e.g., Fig. \ref{fig:arrow-patterns-diamond}(b) when $\lambda=-1$,
$\boldsymbol{\chi}_{i,k}^{(\mathcal{S})}$ is zero and subsystem $\mathcal{S}$
is isolated from the whole network. While when $\boldsymbol{\chi}_{i,k}^{(\mathcal{S})}$
is significant, it indicates that the environment outside the subspace
$\mathcal{S}$ should have considerable impact. $\boldsymbol{\chi}_{i,k}^{(\mathcal{S})}$
usually varies when parameters of the network change. As shown in
Fig. \ref{fig:arrow-patterns-diamond}(e)-(h) when $J_{a}$ increases,
all vectors rotate to a certain extent. If local properties of subspace
are not altered, e.g., internal properties in subspace $c$ are intact
when $J_{a}$ increases, the change within the subspace, $\boldsymbol{\chi}_{3,k}^{(c)}$
and $\boldsymbol{\chi}_{4,k}^{(c)}$, are only induced by the change
of the global network. The distributions also show the dependence
of $\boldsymbol{\chi}_{i,k}^{(\mathcal{S})}$ on $J_{\mathcal{S}}$.
The magnitudes of $\boldsymbol{\chi}_{1,k}^{(a)}$ and $\boldsymbol{\chi}_{2,k}^{(a)}$
within subspace $a$ increase with $J_{\mathcal{S}}$, while lengths
of $\boldsymbol{\chi}_{i,k}^{(\mathcal{S})}$ in other subspaces are
not changed dramatically. In addition, the vectors in Fig. \ref{fig:arrow-patterns-diamond}(a)-(d)
exhibit symmetric distribution along site 1-4-3 and 1-2-3, indicating
the system can be further reduced to a three-site linear chain. While
in Fig. \ref{fig:arrow-patterns-diamond}(e)-(h) when $J_{a}=1.5$
the symmetry of the distribution of $\boldsymbol{\chi}_{i,k}^{(\mathcal{S})}$
is broken and the system is irreducible.

The distribution shown in Fig. \ref{fig:arrow-patterns-diamond} has
extra significance besides the separation of local and global properties.
In Eq. (\ref{eq:coefficient-vector}), $C_{i,k}=\boldsymbol{u}_{i,k}^{(\mathcal{S})}\boldsymbol{\chi}_{i,k}^{(\mathcal{S})}$
allows one to read the component of wave function of each site directly
from the inner product of vectors $\boldsymbol{u}_{i,k}^{(\mathcal{S})}$
and $\boldsymbol{\chi}_{i,k}^{(\mathcal{S})}$. Since the matching
condition assumes the equivalent $C_{i,k}$ shared among subspaces,
we may choose a pair of vectors arbitrarily in any involved subspace
$\mathcal{S}$. The vectors help identify immediately all zero components
that do not contribute to the amplitude of a specific site. In Fig.
\ref{fig:arrow-patterns-diamond}(b), $\boldsymbol{\chi}_{i,2}^{(\mathcal{S})}$
being zero vector leads to $C_{i,2}=0$ for all sites when $\lambda_{2}=-1$.
In Fig. \ref{fig:arrow-patterns-diamond}(c), vectors $\boldsymbol{u}_{3,3}^{(\mathcal{S})}$
and $\boldsymbol{\chi}_{3,3}^{(\mathcal{S})}$ for site 3 being orthogonal
also results in $C_{3,3}=0$. While when $J_{a}=1.5$ as shown in
Fig. \ref{fig:arrow-patterns-diamond}(g), $\boldsymbol{\chi}_{3,3}^{(\mathcal{S})}$
rotates slightly and $C_{3,3}$ is no longer zero due to the breaking
of the orthogonality. 

Thereby, when parameters are altered, the change of all vectors separated
by local and global contributions can be simultaneously traced on
the graph, and the wave functions can be easily determined. It offers
the possibility to design and manipulate the vectors to control the
quantum evolution on the graph for optimized quantum state transfer.

\subsection{Transport efficiency in trimer\label{sub:trimer-efficiency}}

\begin{figure}
\includegraphics[bb=3bp -1bp 2024bp 1500bp,scale=0.11]{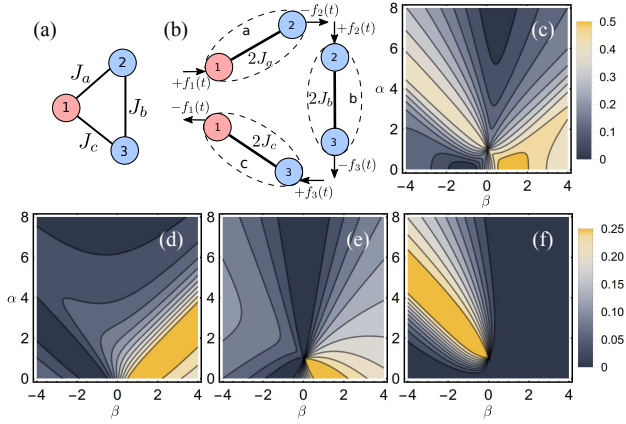}

\caption{The dimerized decomposition scheme. (a) A closed-loop trimer ($K_{3}$
graph) is decomposed into (b) a set of subsystems (dashed circles),
each of which consists of two sites, and subsystems communicate via
flows $f_{i}$ (arrows). The initial excitation starts from site 1
(red), and sites 2 and 3 are targets (blue). Panel (c) shows the efficiency
distribution $\eta_{2}(\beta,\alpha)$. Panels (d)-(f) show the partial
contributions of path $m$ as introduced in Eq. (\ref{eq:ft-of-trimer}).
\label{fig:scheme-trimer-and-eta}}
\end{figure}

In this section, the trimer model is examined with the dimerized decomposition.
The trimer model has the potential application to optimize the excitation
energy transfer via biomolecular network, e.g., the excitation transfer
from B800 to B850 bacteriochlorophylls (BChls) in light-harvesting
complex II (LH2). With the structure dimerization of the B850 ring
\cite{grondelle_energy_2006,yang_dimerization-assisted_2010,scholes_lessons_2011},
a trimer can be viewed as a subunit of the two-layer rings, between
which a carotenoid connects B800 BChl (source) with one of the two
B850 BChls (traps). The excitation transport from a source site to
the two-site traps within the single-exciton manifold can be investigated
with the trimer as shown in Fig. \ref{fig:scheme-trimer-and-eta}(a).

The trimer has the source of the excitation energy at site $1$, and
the target sites 2 and 3 are supposed to trap the energy. For simplicity,
the on-site energies $\varepsilon_{i}$ and decoherence rates $\varGamma_{i}$
are assumed identical for all sites. Coupling strengths among three
sites, $J_{a}$, $J_{b}$ and $J_{c}$, are adjustable, e.g., by changing
spatial distances between the sites. The Hamiltonian for the trimer
system is $\hat{H}=\sum_{i}^{3}\varepsilon_{i}|i\rangle\text{\textlangle}i|+J_{a}|1\rangle\text{\textlangle}2|+J_{b}|2\text{\textrangle}\text{\textlangle}3|+J_{c}|3\text{\textrangle}\text{\textlangle}1|+\text{c.c.}$.
Here we use the dimerized decomposition to calculate the amplitudes.
The decomposition is shown in Fig. \ref{fig:scheme-trimer-and-eta}(b)
with the EOMs of subsystems determined by Eq. (\ref{eq:eom-decomposition-matrix-form}).
The amplitudes $\boldsymbol{c}^{(\mathcal{S})}$ are $\boldsymbol{c}^{(a)}=\left(c_{1}^{(a)},c_{2}^{(a)}\right){}^{\mathrm{T}}$,
$\boldsymbol{c}^{(b)}=\left(c_{2}^{(b)},c_{3}^{(b)}\right){}^{\mathrm{T}}$
and $\boldsymbol{c}^{(c)}=\left(c_{3}^{(c)},c_{1}^{(c)}\right){}^{\mathrm{T}}$.
Note that since $n_{i}=2$ for all sites in the $K_{3}$ graph, the
amplitude in the full system $c_{i}$ is evenly distributed in the
subsystems, $c_{i}^{(\mathcal{S})}=c_{i}/2$. Accordingly, the initial
conditions are $\boldsymbol{c}^{(a)}(0)=\left(1/2,0\right){}^{\mathrm{T}}$,
$\boldsymbol{c}^{(b)}(0)=\left(0,0\right){}^{\mathrm{T}}$ and $\boldsymbol{c}^{(c)}(0)=\left(0,1/2\right){}^{\mathrm{T}}$.
Using the junction rule, it is convenient to directly define $\boldsymbol{f}=\left(\boldsymbol{f}^{(a)},\boldsymbol{f}^{(a)},\boldsymbol{f}^{(c)}\right)^{\mathrm{T}}$
with $\boldsymbol{f}^{(a)}=\left(f_{1},-f_{2}\right){}^{\mathrm{T}}$,
$\boldsymbol{f}^{(b)}=\left(f_{2},-f_{3}\right){}^{\mathrm{T}}$ and
$\boldsymbol{f}^{(c)}=\left(f_{3},-f_{1}\right){}^{\mathrm{T}}$.
Assuming $\varepsilon_{i}=0$, the local Hamiltonian of subsystem
$\mathcal{S}$ is $\boldsymbol{H}^{(\mathcal{S})}=2J_{\mathcal{S}}\begin{pmatrix}0 & 1\\
1 & 0
\end{pmatrix}$. Substituting $\boldsymbol{H}^{(\mathcal{S})}$, $\boldsymbol{c}^{(\mathcal{S})}$
and $\boldsymbol{f}^{(\mathcal{S})}$ into Eq. (\ref{eq:eom-decomposition-matrix-form}),
we obtain the EOMs of all subsystems.

The matching conditions in the $s$-domain, $\tilde{c}_{2}^{(a)}=\tilde{c}{}_{2}^{(b)}$,
$\tilde{c}_{3}^{(b)}=\tilde{c}_{3}^{(c)}$ and $\tilde{c}_{1}^{(c)}=\tilde{c}_{1}^{(a)}$,
yield the linear system with respect to $\tilde{f}_{i}$, \begin{widetext}

\begin{equation}
\begin{pmatrix}-\tilde{u}_{22}^{(c)}-\tilde{u}_{11}^{(a)} & \tilde{u}_{12}^{(a)} & \tilde{u}_{21}^{(c)}\\
\tilde{u}_{21}^{(a)} & -\tilde{u}_{22}^{(a)}-\tilde{u}_{11}^{(b)} & \tilde{u}_{12}^{(b)}\\
\tilde{u}_{12}^{(c)} & \tilde{u}_{21}^{(b)} & -\tilde{u}_{22}^{(b)}-\tilde{u}_{11}^{(c)}
\end{pmatrix}\begin{pmatrix}\tilde{f}_{1}\\
\tilde{f}_{2}\\
\tilde{f}_{3}
\end{pmatrix}=\begin{pmatrix}\frac{1}{n_{1}}[-\tilde{u}_{22}^{(c)}+\tilde{u}_{11}^{(a)}]\\
-\frac{1}{n_{1}}\tilde{u}_{21}^{(a)}\\
\frac{1}{n_{1}}\tilde{u}_{12}^{(c)}
\end{pmatrix},\label{eq:matching-conditions-trimer}
\end{equation}
\end{widetext}where $\tilde{u}_{ij}^{(\mathcal{S})}$ is the element
of $\boldsymbol{\tilde{U}}^{(\mathcal{S})}=\frac{1}{4J_{s}^{2}+s{}^{2}}\begin{pmatrix}s & -2iJ_{\mathcal{S}}\\
-2iJ_{\mathcal{S}} & s
\end{pmatrix}$ reduced from Eq. (\ref{eq:Us}). Solving Eq. (\ref{eq:matching-conditions-trimer})
yields 

\begin{eqnarray}
\tilde{f}_{1}(s) & = & \frac{s(J_{a}^{2}-J_{c}^{2})}{g(s)},\nonumber \\
\tilde{f}_{2}(s) & = & \frac{sJ_{c}J_{b}-iJ_{a}(s^{2}+2J_{b}^{2})}{g(s)},\nonumber \\
\tilde{f}_{3}(s) & = & -\frac{sJ_{a}J_{b}-iJ_{c}(s^{2}+2J_{b}^{2})}{g(s)}\label{eq:fs-of-trimer}
\end{eqnarray}
with $g(s)=2[s^{3}+(J_{a}^{2}+J_{b}^{2}+J_{c}^{2})s-2iJ_{a}J_{b}J_{c}]$
the characteristic polynomial. Note that $\tilde{f}_{2}(s)$ and $\tilde{f}_{3}(s)$
differ by exchanging subscripts $a$ and $c$ since they are treated
on the equal footing within subsystem $b$. The signs of $\tilde{f}_{2}(s)$
and $\tilde{f}_{3}(s)$ also differ due to the definitions as the
flow-in and flow-out, respectively. By applying the inverse Laplace
transform, we obtain
\begin{eqnarray}
f_{1}(t) & = & -\frac{i}{2}\sum_{m}\frac{(J_{a}^{2}-J_{c}^{2})\lambda_{m_{1}}}{(\lambda_{m_{1}}-\lambda_{m_{2}})(\lambda_{m_{1}}-\lambda_{m_{3}})}e^{i\lambda_{m_{1}}t},\nonumber \\
f_{2}(t) & = & -\frac{i}{2}\sum_{m}\frac{J_{b}J_{c}\lambda_{m_{1}}+J_{a}(\lambda_{m_{1}}^{2}-2J_{b}^{2})}{(\lambda_{m_{1}}-\lambda_{m_{2}})(\lambda_{m_{1}}-\lambda_{m_{3}})}e^{i\lambda_{m_{1}}t},\label{eq:ft-of-trimer}
\end{eqnarray}
where the integer index $m_{i}$ is the $i$th member of $m$, the
member of a cyclically ordered set, i.e., $(m_{1},m_{2},m_{3})\in\{(1,2,3),(2,3,1),(3,1,2)\}$.
$\lambda_{j}$ are zeros of $g(i\lambda)$ whose solutions are $\lambda_{j}\in2\kappa\{-\cos(\theta/3),\cos[(\theta-\pi)/3],\cos[(\theta+\pi)/3]\}$
with $\kappa=[(J_{a}^{2}+J_{b}^{2}+J_{c}^{2})/3]^{1/2}$ and $\theta=\arg[J_{a}J_{b}J_{c}+i[\kappa^{6}-(J_{a}J_{b}J_{c})^{2}]^{1/2}]$.
$f_{3}(t)$ can also be obtained by exchanging subscripts $a$ and
$c$ in $-f_{2}(t)$. It is shown that the zeros satisfy $\sum_{j}\lambda_{j}=0$.
Moreover, $\theta=\arctan[(J_{a}^{2}+J_{b}^{2}+J_{c}^{2})^{3}/(3^{3}J_{a}^{2}J_{b}^{2}J_{c}^{2})-1]^{1/2}$
with $\kappa^{6}-(J_{a}J_{b}J_{c})^{2}\ge0$. Breaking any edge, e.g.,
$J_{a}=0$, results in $\theta=\pi/2$ and $\lambda_{j}\in\{-\sqrt{3},\sqrt{3},0\}\kappa$.
In addition, if $J_{a}^{2}=J_{b}^{2}=J_{c}^{2}$, we have $\theta=0$
and degeneracy occurs as $\lambda_{j}\in\{-2,1,1\}\kappa$.

Substituting $\boldsymbol{\tilde{U}}^{(\mathcal{S})}(s)$, $\boldsymbol{\tilde{f}}^{(\mathcal{S})}(s)$
and $\boldsymbol{c}^{(\mathcal{S})}(0)$ into Eq. (\ref{eq:cs-of-local-eom})
we find $\tilde{c}_{1}(s)=(s^{2}+J_{b}^{2})/g(s)$ and $\tilde{c}_{2}(s)=(isJ_{a}+J_{b}J_{c})/g(s)$.
The amplitude $\tilde{c}_{3}(s)$ is similar to $\tilde{c}_{2}(s)$
differing by exchanging $a$ and $c$. Back to the time domain, the
amplitudes of sites $1$ and $2$ are $c_{i}(t)=\frac{1}{2}\sum_{m}C_{i}^{(m)}e^{i\lambda_{m_{1}}t},(i=1,2)$
with 
\begin{eqnarray}
C_{1}^{(m)} & = & \frac{\lambda_{m_{1}}^{2}-J_{b}^{2}}{(\lambda_{m_{1}}-\lambda_{m_{2}})(\lambda_{m_{1}}-\lambda_{m_{3}})},\nonumber \\
C_{2}^{(m)} & = & \frac{J_{a}\lambda_{m_{1}}-J_{b}J_{c}}{(\lambda_{m_{1}}-\lambda_{m_{2}})(\lambda_{m_{1}}-\lambda_{m_{3}})}.\label{eq:ct-of-trimer}
\end{eqnarray}
The above derivation assumes that $\bar{\varepsilon}=0$. If the decoherence
rate $\varGamma$ is considered for each site \cite{mulken_survival_2007},
$\bar{\varepsilon}=-i\frac{\varGamma}{2}$. Accordingly, $s\rightarrow s+\frac{\varGamma}{2}$
in Eq. (\ref{eq:fs-of-trimer}), and in $f_{i}(t)$ and $c_{i}(t)$,
$e^{i\lambda_{k}t}\rightarrow e^{i\lambda_{k}t-\varGamma t/2}$.

We apply the method to calculate the excitation transfer efficiency
$\eta_{i}$ toward site $i$ and its dependence on $J_{\mathcal{S}}$
and $\varGamma$. The efficiency is defined by $\eta_{i}=\lim_{t\rightarrow\infty}[\sigma_{i}(t)/\sum_{j}\sigma_{j}(t)]$
with $\sigma_{i}(t)=\int_{0}^{t}d\tau|c_{i}(\tau)|^{2}$ the accumulated
population trapped at site $i$ by time $t$. With the denominator
$\sum_{j}\sigma_{j}(\infty)=1/\varGamma$, we have $\eta_{i}=\varGamma\sigma_{i}(\infty)$.
Substituting $c_{i}(t)=n_{i}c_{i}^{(\mathcal{S})}(t)=\sum_{m}C_{i}^{(m)}e^{i\lambda_{m_{1}}t-\varGamma t/2}$
into $\eta_{i}$, we find $\eta_{i}=\varGamma\sum_{m,n}(C_{i}^{(m)})^{*}C_{i}^{(n)}/[i(\lambda_{m_{1}}-\lambda_{n_{1}})+\varGamma]$.
With $\lambda_{m_{1}},C_{i}^{(m)}\in\mathbb{R}$, the efficiency
is given by  
\begin{equation}
\eta_{i}=\sum_{m}|C_{i}^{(m)}|^{2}+2\sum_{m<n}\frac{C_{i}^{(m)}C_{i}^{(n)}}{1+[(\lambda_{m_{1}}-\lambda_{n_{1}})/\varGamma]^{2}},\label{eq:excitation-transfer-efficiency}
\end{equation}
with the first non-interfering sum and the last interfering part.
The ratio between $(\lambda_{m_{1}}-\lambda_{n_{1}})$ and $\varGamma$
decides the contribution of the interfering part to $\eta_{i}$. When
$\varGamma$ is small, the interfering term vanishes and $\eta_{i}\rightarrow\sum_{m}|C_{i}^{(m)}|^{2}$.
On the other hand, the contribution of interference increases with
$\varGamma$. When $\varGamma\rightarrow\infty$, $\eta_{i}=\sum_{m<n}|C_{i}^{(m)}+C_{i}^{(n)}|^{2}$
approaches the limit when the decoherence induced destructive interference
dominates and $\eta_{2}$ is low in general.

The efficiency of excitation transfer toward site 2, $\eta_{2}$,
is evaluated by substituting $C_{2}^{(m)}$ in Eq. (\ref{eq:ct-of-trimer})
into Eq. (\ref{eq:excitation-transfer-efficiency}). Alternatively,
$\eta_{i}$ can be analyzed using $\tilde{c}_{i}(s)$ in the $s$-domain
without the necessity of the inverse Laplace transform back to the
time domain, because the relevant information is embedded in poles
of $m$, and the $C_{i}^{(m)}$ is exactly the residues of $\tilde{c}_{i}(s)$
in the complex $s$-plane.

By introducing two dimensionless parameters $\alpha$ and $\beta$,
we define $J_{a}=(1+\beta)J$, $J_{c}=(1-\beta)J$ and $J_{b}=\alpha J$.
The parameter $\beta$ describes the asymmetry for the upper and lower
source-trap couplings, and $\alpha$ accounts for the inter-trap coupling.
In the LH2 complex, $\alpha$ characterizes the dimerization of the
B850 BChl ring that tunes the coupling $J_{b}$ between neighboring
B850 BChls, and $\beta$, as the difference between $J_{a}$ and $J_{c}$,
describes the spatial deformation when rotating the B850 ring relative
to the B800 ring \cite{yang_dimerization-assisted_2010}.

The efficiency $\eta_{2}(\beta,\alpha)$ is shown in Fig. \ref{fig:scheme-trimer-and-eta}(b)
for $\varGamma=0.01J$  with the distinct negative- and positive-$\beta$
distributions. In the negative-$\beta$ region, the excitation is
transferred mostly via the indirect path $1\leftrightarrow3\leftrightarrow2$,
whereas in the positive-$\beta$ region the direct path $1\leftrightarrow2$
dominates the contribution. Changing the positivity of $\beta$ adjusts
the ratio of contributions from direct and indirect paths. The global
maximum of $\eta_{2}$ present in the positive-$\beta$ region suggests
the transfer of high efficiency should favor the direct path.

When $\beta=1$, the indirect path is completely blocked. At $(\beta,\alpha)=(1,0)$,
the maximal $\eta_{2}=1/2$ is obtained and the excitation is transferred
via $1\leftrightarrow2$ directly. When $\beta=-1$, the direct path
is blocked instead and the transfer depends only on the indirect path
$1\leftrightarrow3\leftrightarrow2$. Particularly when $\alpha=0$,
site 2 is isolated and $\eta_{2}=0$. Either via direct or indirect
path, the high-$\eta_{2}$ distribution is roughly along $\beta=1+|\alpha|$
when $\varGamma\ll J$. 

When $\beta=0$, the identical paths $1\leftrightarrow2$ and $1\leftrightarrow3$
render sites $2$ and $3$ a single effective site with the $\alpha$-tunable
energy levels. When $\alpha=0$, the energy of the effective site
is $\varepsilon_{1}$. Increasing $\alpha$, however, the energy level
splits leading to an increasingly large energy gap with the center
moving away from $\varepsilon_{1}$ that lowers $\eta_{2}$. When
the system is symmetric as $J_{a}=J_{b}=J_{c}$, i.e., $(\beta,\alpha)=(0,1)$,
the degeneracy occurs and $\eta_{2}=1/3$.

Since $\eta_{2}$ depends almost on the non-interfering part when
$\varGamma<J$, we show in Fig. \ref{fig:scheme-trimer-and-eta}(d)-(f)
the partial contributions from $|C^{(m)}|^{2}$. Roughly, $m=(1,2,3)$
and $(2,3,1)$ correspond to the direct paths and $m=(3,1,2)$ the
indirect path. Each path contributes $\sim1/4$ to the efficiency,
and the maximum $\eta_{2}=1/2$ is when the partial contributions
of direct paths, Fig. \ref{fig:scheme-trimer-and-eta}(d) and (e),
overlap at $(\beta,\alpha)=(1,0)$.

\begin{figure}
\includegraphics[scale=0.11]{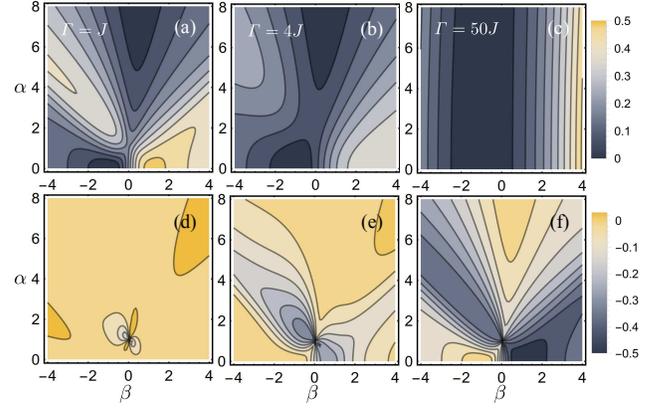}\caption{Efficiencies $\eta_{2}(\beta,\alpha)$ for $\varGamma=J$, $4J$ and
$50J$ are shown in panels (a)-(c), respectively. In panel (c) for
$\varGamma=50J$, the value of $\eta_{2}$ is low and hence multiplied
by a factor $25$ to highlight the distribution. The corresponding
partial contributions from the interfering part in Eq. (\ref{eq:excitation-transfer-efficiency})
are shown in panels (d)-(f). \label{fig:efficiency-vs-gamma}}
\end{figure}

Fig. \ref{fig:efficiency-vs-gamma}(a) confirms that $\eta_{2}$ is
dominated by the non-interfering sum of $|C_{2}^{(m)}|^{2}$ when
$\varGamma<J$, as depicted in Eq. (\ref{eq:excitation-transfer-efficiency}),
which also indicates the non-interfering part is $\varGamma$ irrelevant,
as shown by the similar patterns in Fig. \ref{fig:scheme-trimer-and-eta}(c)
and Fig. \ref{fig:efficiency-vs-gamma}(a). In Fig. \ref{fig:efficiency-vs-gamma}(c)
when $\varGamma\gg J$, however, the interfering part plays an increasingly
important role and $\eta_{2}$ becomes $\varGamma$-dependent. Increasing
$\varGamma$, the maximum of $\eta_{2}$, which is located at $(\beta,\alpha)=(1,0)$
when $\varGamma<J$, moves outward along the $\beta$-axis. It suggests
that the efficiency is deteriorated by dissipations on all sites and
a stronger coupling $J_{b}$ is in need for the maximized efficiency.
The origin of the low $\eta_{2}$ when $\varGamma\gg J$ is due
to the negative contribution from the interfering part, as shown in
Fig. \ref{fig:efficiency-vs-gamma}(d)-(f). Being negative, the interfering
part intensifies with increasing $\varGamma$ until its distribution
resembles the non-interfering part, which is neutralized by the former
yielding the low $\eta_{2}$.

In summary, we have introduced the method of dimerized decomposition
to study the quantum evolution on a graph. The method allows for the
separation of the local subsystems from the global network and offers
insights from the perspective of perturbing flows among sites. The
decomposition is applied to a diamond graph for demonstration, and
the EOMs can be easily generated using the diagrammatic technique.
The method allows for observing the distribution of vectors representing
influences from the local time evolution and the perturbation from
the global network. Furthermore, we apply the method to analyze the
model of a source-trap-trap trimer, on which the excitation transfer
efficiency influenced by the symmetry of source-trap couplings and
the inter-trap coupling is investigated. With contributions from direct
and indirect paths separated, the transfer efficiency is deteriorated
by decoherence-induced destructive interference. Beyond examples we
have shown in this work, the dimerized decomposition is universal
and straightforward for further extensions towards arbitrary graphs.
Besides applications to study the transport efficiency, as the graph
presented here mapping the state-to-state transition, the relation
between the local subsystem and global network may provide the measurement
to better understand concepts like entanglement in multi-particle
quantum systems.

\section*{Appendix\label{sec:Appendix}}

In Sec. \ref{sub:diamond_graph} for the diamond graph, the flow functions
need to be determined by solving the linear system $\tilde{\boldsymbol{M}}\tilde{\boldsymbol{f}}=\tilde{\boldsymbol{b}}$.
Given flow functions of the order $\tilde{\boldsymbol{f}}=(\tilde{f}_{1}^{(a)},\tilde{f}_{2}^{(a)},\tilde{f}_{2}^{(b)},\tilde{f}_{3}^{(b)},\tilde{f}_{3}^{(c)},\tilde{f}_{4}^{(c)},\tilde{f}_{4}^{(d)},\tilde{f}_{1}^{(d)},\tilde{f}_{2}^{(e)},\tilde{f}_{4}^{(e)})^{\mathrm{T}}$,
the $10\times10$ matrix $\tilde{\boldsymbol{M}}$ constructed from
both the junction rules and the matching conditions reads

\begin{widetext}
\begin{eqnarray*}
\tilde{\boldsymbol{M}} & = & \begin{pmatrix}\tilde{u}_{1,1}^{(a)} & \tilde{u}_{1,2}^{(a)} & 0 & 0 & 0 & 0 & -\tilde{u}_{2,1}^{(d)} & -\tilde{u}_{2,2}^{(d)} & 0 & 0\\
\tilde{u}_{2,1}^{(a)} & \tilde{u}_{2,2}^{(a)} & -\tilde{u}_{1,1}^{(b)} & -\tilde{u}_{1,2}^{(b)} & 0 & 0 & 0 & 0 & 0 & 0\\
\tilde{u}_{2,1}^{(a)} & \tilde{u}_{2,2}^{(a)} & 0 & 0 & 0 & 0 & 0 & 0 & -\tilde{u}_{1,1}^{(e)} & -\tilde{u}_{1,2}^{(e)}\\
0 & 0 & \tilde{u}_{2,1}^{(b)} & \tilde{u}_{2,2}^{(b)} & -\tilde{u}_{1,1}^{(c)} & -\tilde{u}_{1,2}^{(c)} & 0 & 0 & 0 & 0\\
0 & 0 & 0 & 0 & \tilde{u}_{2,1}^{(c)} & \tilde{u}_{2,2}^{(c)} & -\tilde{u}_{1,1}^{(d)} & -\tilde{u}_{1,2}^{(d)} & 0 & 0\\
0 & 0 & 0 & 0 & \tilde{u}_{2,1}^{(c)} & \tilde{u}_{2,2}^{(c)} & 0 & 0 & -\tilde{u}_{2,1}^{(e)} & -\tilde{u}_{2,2}^{(e)}\\
{\color{blue}1} & {\color{blue}0} & {\color{blue}0} & {\color{blue}0} & {\color{blue}0} & {\color{blue}0} & {\color{blue}0} & {\color{blue}1} & {\color{blue}0} & {\color{blue}0}\\
{\color{blue}0} & {\color{blue}1} & {\color{blue}1} & {\color{blue}0} & {\color{blue}0} & {\color{blue}0} & {\color{blue}0} & {\color{blue}0} & {\color{blue}1} & {\color{blue}0}\\
{\color{blue}0} & {\color{blue}0} & {\color{blue}0} & {\color{blue}1} & {\color{blue}1} & {\color{blue}0} & {\color{blue}0} & {\color{blue}0} & {\color{blue}0} & {\color{blue}0}\\
{\color{blue}0} & {\color{blue}0} & {\color{blue}0} & {\color{blue}0} & {\color{blue}0} & {\color{blue}1} & {\color{blue}1} & {\color{blue}0} & {\color{blue}0} & {\color{blue}1}
\end{pmatrix}.
\end{eqnarray*}
\end{widetext}

The first six rows are from matching conditions (\ref{eq:matching-condition}) and the rest are from the junctions rules (\ref{eq:junction-rule}).
The array of initial conditions is given by 
\begin{eqnarray*}
\tilde{\boldsymbol{b}} & = & \begin{pmatrix}(\tilde{u}_{1,1}^{(a)}-u_{2,2}^{(d)})c_{1}(0)+\tilde{u}_{1,2}^{(a)}c_{2}(0)-\tilde{u}_{2,1}^{(d)}c_{4}(0)\\
\tilde{u}_{2,1}^{(a)}c_{1}(0)+(\tilde{u}_{2,2}^{(a)}-\tilde{u}_{1,1}^{(b)})c_{2}(0)-\tilde{u}_{1,2}^{(b)}c_{3}(0)\\
\tilde{u}_{2,1}^{(a)}c_{1}(0)+(\tilde{u}_{2,2}^{(a)}-\tilde{u}_{1,1}^{(e)})c_{2}(0)-\tilde{u}_{1,2}^{(e)}c_{4}(0)\\
\tilde{u}_{2,1}^{(b)}c_{2}(0)+(\tilde{u}_{2,2}^{(b)}-\tilde{u}_{1,1}^{(c)})c_{3}(0)-\tilde{u}_{1,2}^{(c)}c_{4}(0)\\
-\tilde{u}_{1,2}^{(d)}c_{1}(0)+\tilde{u}_{2,1}^{(c)}c_{3}(0)+(\tilde{u}_{2,2}^{(c)}-\tilde{u}_{1,1}^{(d)})c_{4}(0)\\
-\tilde{u}_{2,1}^{(e)}c_{2}(0)+\tilde{u}_{2,1}^{(c)}c_{3}(0)+(\tilde{u}_{2,2}^{(c)}-\tilde{u}_{2,2}^{(e)})c_{4}(0)\\
{\color{blue}0}\\
{\color{blue}0}\\
{\color{blue}0}\\
{\color{blue}0}
\end{pmatrix}.
\end{eqnarray*}

Since the initial amplitude $c_{i}(0)$ distributed in different subsystems
for site $i$ are the same, here label $\mathcal{S}$ to indicate
specific subsystem is neglected, and hence $c_{i}(0)$ needs to be
substituted by the full amplitude $c_{i}(0)$ divided by corresponding
connectivity $n_{i}$. 

\begin{acknowledgments}
This work is supported by Shanghai Sailing Program (16YF1412600);
National Basic Research Program of China (2013CB922200); the National
Natural Science Foundation of China (11420101003, 11604347, 91636105).
T.-M. Yan thanks M. Weidem\"{u}ller for remarks and suggestions.
\end{acknowledgments}

\bibliographystyle{apsrev4-1}
\bibliography{dimerized_decomposition}

\end{document}